\documentclass[12pt]{article}
\usepackage{url} 
\usepackage{amsmath}
\usepackage{amssymb}
\usepackage{graphicx}
\usepackage{float}
\usepackage{soul}
\usepackage{siunitx}
\usepackage{authblk}
\usepackage{hyperref}
\usepackage{enumerate}
\usepackage[margin=1in]{geometry}

\title{Sealing Europa's vents by vapor deposition:\\ An order of magnitude study}

\author[1]{S. Boccelli \thanks{stefano.boccelli@nasa.gov}}
\author[2]{S. R. Carberry Mogan}
\author[3,4]{R. E. Johnson}
\author[1]{O. J. Tucker}
\affil[1]{\small NASA Goddard Space Flight Center, USA}
\affil[2]{\small University of California, Berkeley, USA}
\affil[3]{\small University of Virginia, Wilsdorf Hall, USA}
\affil[4]{\small New York University, USA}

\begin{document}

\maketitle

\begin{abstract}
Fractures and vents in the ice crust of Europa, exposing the sub-surface ocean or liquid-water inclusions to the vacuum, 
might be responsible for the generation of water-vapor plumes.
During its passage through the ice, the plume vapor is expected to partially condense on the cold ice walls. 
Together with other effects (water spillage, compression forces, etc.) this mechanism likely contributes to sealing the vent. 
In this work, we develop a simple lumped-parameter model that can quantify how quickly a hypothetical vent of prescribed width 
would be sealed via water-vapor deposition.
As an example, we apply our model to the vent size and density conditions inferred from the 2012 Hubble Space Telescope plume detection,
predicting a sealing time of about 30 minutes.
This suggests that the actual ice fracture might have been larger than originally proposed and/or the plume density at the vent might have been lower.
While many other effects could have been present and responsible for sealing the vent, 
our estimates indicate that vapor deposition might have played a major role in eventually shutting off the observed plume.
A map of sealing times vs. plume density, mass flow rate and aperture areas is given.
Plume quantities from the literature are analyzed and compared to our results. 
For a given plume density/mass flow rate, small apertures would be sealed quickly by vapor deposition and are thus incompatible with observations.
\end{abstract}

%
%
  
\section{Introduction}

Fractures and vents in the surface of icy moons, exposing the sub-surface ocean or liquid-water inclusions to the low-pressure atmosphere, generate water-vapor plumes that expand into space.
Such plumes are routinely observed on Enceladus \cite{waite2006cassini,hansen2006enceladus, hedman2009spectral,villanueva2023jwst} 
and work is underway to reproduce them in the laboratory \cite{bourgeois2024pisces,bourgeois2024leaky}.
On Europa, these plumes are more elusive and the few confirmed observations \cite{roth2014transient,jia2018evidence,paganini2020measurement}
are complemented with both acclaimed but questioned detections and non-detections \cite{huybrighs2020active,jia2021comment,huybrighs2021reply,sparks2016probing,giono2020analysis,villanueva2023endogenous,hansen2024juno,cordiner2024alma,kimura2024search}.

Topography maps of Europa \cite{greeley2000geologic,quick2020characterizing} suggest that plumes might generate from a variety of geologic formations.
These include two-dimensional fractures spanning a large portion of the surface \cite{roth2014transient}.
However, in most of the literature concerning in-situ observations or numerical plume modeling, vapor plumes are assumed to originate from a cylindrical or from a point source \cite{
teolis2017plume,arnold2019magnetic,huybrighs2020active,paganini2020measurement,vorburger2021modeling,tseng20223d}. 
This assumption is often made for simplicity, but can be physically relevant for plumes originating from near-surface water inclusions and diapirs \cite{vorburger2021modeling} or from localized surface features that could be evidence of cryovolcanism \cite{noviello2019mapping}.
The upcoming imaging, particle and field, and thermal measurements of the Europa Clipper \cite{vance2023investigating} and JUpiter ICy moons Explorer (Juice) \cite{grasset2013jupiter} missions will help to shed light on the geologic origin of possible plumes and attempt to probe their composition \cite{winterhalder2022assessing,teolis2017plume}.
Developing theoretical and numerical models of the different physical processes associated with water-vapor plumes is crucial to inform the missions.

The lifetime of plumes is related to the time at which the apertures in the surface remain open \cite{quick2013constraints}.
These apertures can be sealed via various mechanisms, including deformation of the ice shell induced by the subsurface ocean, 
water spillage and freezing, nucleation of ice grains created within a possible plume and their deposition on the walls \cite{spencer2018plume,vincent2022numerical,schmidt2008slow}. 
In this work, we investigate whether molecular deposition on the walls of the aperture can have 
a significant effect on the sealing process, and we analyze its characteristic time scales.
Our results provide order-of-magnitude bounds for the possible lifetime of a plume.
For a given plume density, we show that vapor deposition alone would seal a sufficiently small vent within a short time. 
This poses limitations on the vent sizes commonly inferred from observations and will provide additional guidelines for numerical plume simulations. 

While we do not necessarily claim that vapor deposition alone is responsible for explaining the elusive nature of Europa's plumes, we show that, 
under certain vent-size conditions, this can become a significant---if not dominating---factor.
Analogous conclusions have been drawn from numerical (quasi) one-dimensional simulations of vents on Enceladus \cite{nakajima2016controlled,van2024linking,ingersoll2010subsurface}.
However, it has been proposed that the apertures in Enceladus' south pole are kept open by Saturn's tidal stresses \cite{kite2016sustained}.

\section{Model}\label{sec:model}

\begin{figure}[htpb]
  \centering
  \includegraphics[width=0.7\textwidth]{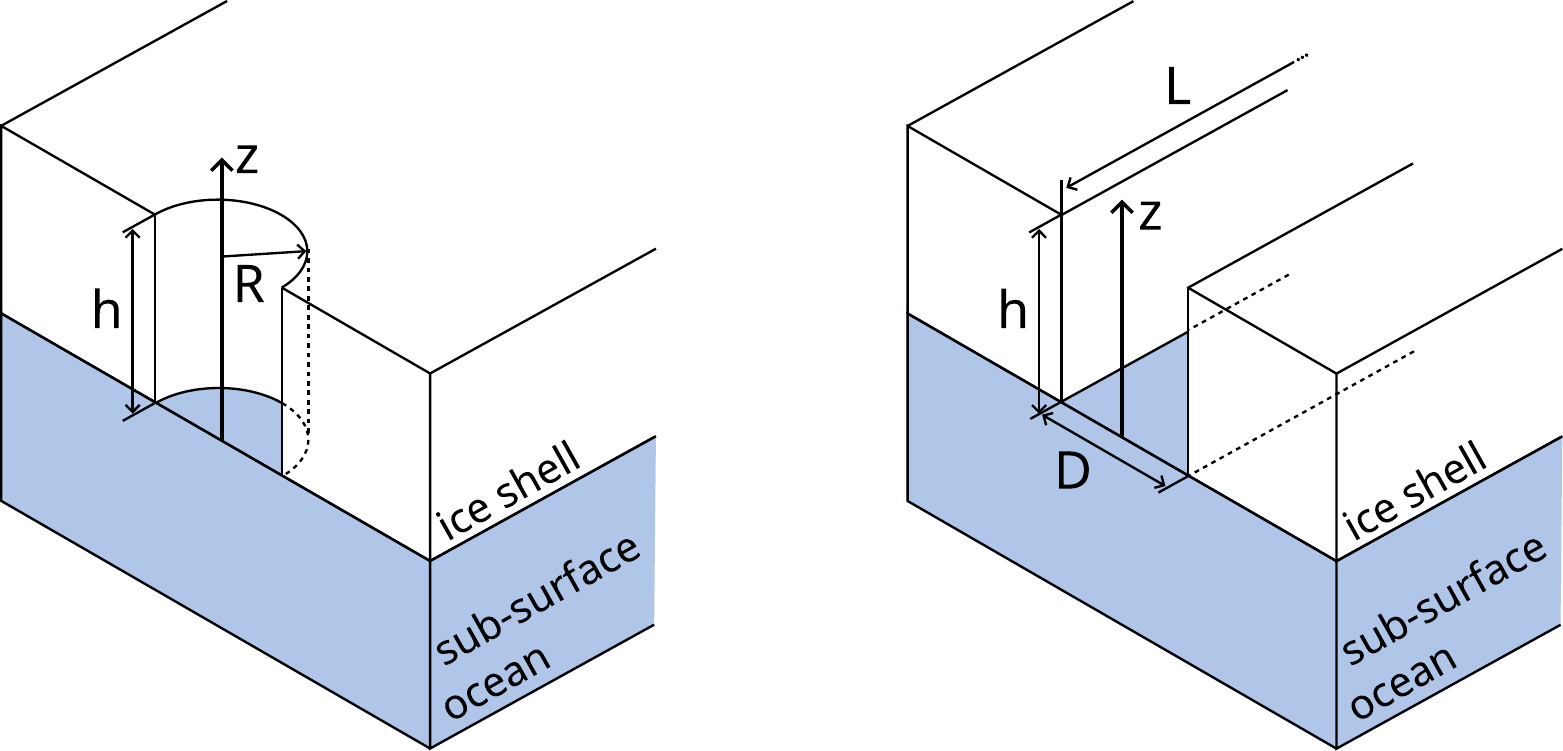}
  \caption{Diagram of the model employed in this work. A plume is created via outgassing from the sub-surface ocean and expands in the direction $z$ through an aperture in the ice shell. Left: cylindrical plume source. Right: two-dimensional fracture-like source.} 
  \label{fig:schematic-plume-condensation}
\end{figure}

We consider a column of water vapor, expanding from an idealized sub-surface liquid-water source to the surface through an aperture as shown in Fig.~\ref{fig:schematic-plume-condensation}-Left.
Europa's tenuous atmosphere is treated as a vacuum.
The liquid-water source might be representative of either a sub-surface ocean or a water inclusion within the ice shell.
We do not consider any depth dependence of the ice temperature, nor the suggested non-uniform nature of the ductile ice-ocean interface \cite{roberts2023exploring}. 
Moreover, in this model we neglect any water spillage or tidal pumping through the ice \cite{spencer2018plume}.
We believe this latter assumption to be reasonable for vents that are small with respect to the local ice thickness, but is most likely inadequate for larger vents and fractures.
In addition, we neglect the cooling of the liquid ocean caused by evaporation, although this might lead to the freezing of the liquid-gas interface, 
depending on the salinity level \cite{ingersoll2016controlled}.
In a real plume we expect all these effects to play a role simultaneously.
Instead, our study aims at isolating the effect of water vapor deposition, to understand its importance.

A plume is assumed to be formed by the pressure differential between the liquid source and the outer vacuum, causing evaporation \cite{nakajima2016controlled}.
Delving into the specific vapor generation mechanism is not necessary for this work, and we prescribe the density of the plume and/or its mass flow rate. 
For simplicity, the number density ($n$), temperature ($T$), and axial bulk velocity ($u$) of the vapor flow are assumed to be uniform within the aperture and constant in time, from the inception to the extinguishing of the plume.
Also, we take $u$ to be equal to the thermal velocity of a Maxwellian distribution, $v_\mathrm{th} = \sqrt{8 k_B T/(\pi m)}$, 
where $k_B$ is the Boltzmann constant and $m$ is the mass of a water molecule. 

A precise estimation of $u$ would require a spatially-resolved numerical analysis and should include the effect of the viscosity and local channel area 
\cite{ingersoll2010subsurface,tucker20152d}.
In our zero-dimensional model (no spacial variations) we lump this parameter into a single value, and we take the thermal velocity as an order-of-magnitude approximation.
This assumption might be not too far from actual values, considering that ice accretion was previously simulated to be maximum near a vent throat, where the flow is
assumed to choke \cite{van2024linking}.

Due to thermal motion, water molecules forming the plume impinge on the lateral surface of the aperture, with a Maxwellian number flux $\Gamma = \tfrac{1}{4} n v_\mathrm{th}$.
Of these, a fraction $\alpha\in[0,1]$ is expected to stick to the surface.
The value of $\alpha$ is related to the residence time and depends on the ice temperature (see \cite{carberry2022callisto} and references therein).
In our model, these molecules build up on the wall, and eventually seal off the vent.

\subsection{Cylindrical vent}

For a cylindrical aperture (Fig.~\ref{fig:schematic-plume-condensation}-Left), we denote the radius by $R$ and compute the axial mass flow rate, generated at the exposed liquid surface, as $\dot{M} = m n \, u \, \pi R^2$.
The number of particles sticking to the vent walls during an infinitesimal time interval, $\mathrm{d} t$, is 
the product of the lateral area by the particle flux, weighted by the sticking coefficient,
\begin{equation}\label{eq:number-molecules-depositing-dt}
  \mathrm{d}N(t) = 2 \pi R(t) h\,  \alpha \Gamma \, \mathrm{d} t \, .
\end{equation}

Here, we denote the ice-shell depth by $h$, but we stress that this parameter eventually cancels out and does not play a role in the model as indicated in the final results.
The particles sticking to the wall gradually fill up the aperture, reducing the radius, $R(t)$, from its initial value, $R_0$.
As they deposit on the wall, the particles create an ice shell of thickness $\mathrm{d} R$.
The number of water molecules within this ice shell is
\begin{equation}\label{eq:N-molecules-shell-of-ice}
  \mathrm{d}N = \frac{\rho_\mathrm{ice}}{m} \, h \, 2 \pi R \mathrm{d} R \, .
\end{equation}

\noindent where $\rho_\mathrm{ice}$ is the mass density of the ice.
From a mass balance at the interface, the sum of Eqs.~\eqref{eq:number-molecules-depositing-dt} and 
 \eqref{eq:N-molecules-shell-of-ice} must be zero, ultimately giving
\begin{equation}\label{eq:dRdt}
  \frac{\mathrm{d} R}{\mathrm{d} t} = - \alpha \Gamma \frac{m}{\rho_{\mathrm{ice}}} \, ,
\end{equation}

\noindent and by integration, the time, $\tau_{\mathrm{cyl}}$, at which the cylindrical aperture closes completely ($R = 0$) is $\tau_{\mathrm{cyl}} = R_0 \rho_\mathrm{ice}/(\alpha \Gamma m)$. 
Expressing it in terms of the number density, $n$, or the \textit{initial} axial mass flow rate, $\dot{M}_0$, one has
\begin{equation}\label{eq:closing-time-cyl}
  \tau_{\mathrm{cyl}} \ = \  
  \frac{4 R_0}{\alpha n} \sqrt{\frac{\pi m}{8 k_B T}}\frac{\rho_\mathrm{ice}}{m}  
  \ = \ 
  \frac{4 \, \rho_\mathrm{ice}}{\alpha \sqrt{\pi}} \frac{A_0^{3/2}}{\dot{M}_0} 
  \, ,
\end{equation}

\noindent where $A_0 = \pi R_0^2$ is the initial cross-sectional area, the initial mass flow rate is $\dot{M}_0 = m n \, u \, A_0$, and where we assumed $u\approx v_\mathrm{th}$ for simplicity.
For a given vapor temperature, $T$, the sealing time depends on the ratio $R_0/n$.
As can be seen in Eq.~\eqref{eq:closing-time-cyl}, $\tau_{\mathrm{cyl}}$ is proportional to $A_0^{3/2}/\dot{M}_0$.
Contours of $\tau_{\mathrm{cyl}}$ are shown in Fig.~\ref{fig:results-literature-axi-plume}.
Also, the ice density affects the sealing times, as it determines the packing of the deposited molecules.
Here, we consider a constant ice density for simplicity, but more accurate models might take it as a function of temperature or of the ice phase.

Equation~\eqref{eq:closing-time-cyl} gives the sealing time as a function of the initial vent quantities.
The time evolution of the mass flow rate that leaves the vent and is injected into the exosphere can be obtained from the solution of Eq.~\eqref{eq:dRdt}, and $\dot{M}(t)$ is found to decrease quadratically in time, 
\begin{equation}
  \dot{M}(t) = m n \, u  \, \pi R(t)^2 = m n \, u \, \pi \left[ R_0 - (\alpha \Gamma m /\rho_\mathrm{ice}) t \right]^2 \, , \ \ \mathrm{for} \ \ t\in[0,\tau_\mathrm{cyl}] \, .
\end{equation}

\subsection{Two-dimensional vent}

The derivation for the two-dimensional case is analogous.
We consider an elongated fracture, depicted in Fig.~\ref{fig:schematic-plume-condensation}-Right, where $L$ is the length, $D_0$ is the initial thickness of the vent, and we assume $L\gg D_0$, as also assumed in \cite{roth2014transient}.
While in the cylindrical case the plume was escaping the ice from a circular vent, here the escape section is rectangular and has an initial area $A_0 = D_0\, L$.
The time required to seal the vent is obtained as a function of the (initial) mass flow rate per unit length, $\dot{M}_0/L$: 
\begin{equation}\label{eq:closing-time-2D}
   \tau_\mathrm{2D} 
= \frac{2 D_0}{\alpha n} \sqrt{\frac{\pi m}{8 k_B T}} \frac{\rho_\mathrm{ice} }{m}
= \frac{2 \rho_\mathrm{ice} D_0^2}{\alpha \left( \dot{M}_0 / L \right)} \, ,
\end{equation}

\noindent and the mass flow rate evolves linearly in time,
\begin{equation} 
  \dot{M}(t) = m n \, u \, L D(t) = m n \, u \, L \left[D_0 - 2 \left(\alpha \Gamma m/\rho_\mathrm{ice} \right) t \right] \ , \ \   \mathrm{for} \ \ t \in [0,\tau_{\mathrm{2D}}] \, .
\end{equation}

\subsection{Caveats: geometric and energy considerations}

Besides the assumptions mentioned above, our model implies a number of further simplifications regarding a possible non-uniform deposition of particles on the walls and the corresponding energy flux.
For quasi-1D numerical models that include a number of these considerations and additional factors, in the context of Enceladus, the reader is referred to \cite{nakajima2016controlled} and \cite{van2024linking}.

First of all, wall deposition constitutes a loss term to the axial mass flow rate equation. 
In other words, only a fraction of the molecules injected into the aperture make it to the exit.
For a large vent this fraction might be negligible at first, but becomes progressively more important as the vapor builds up and the vent size decreases.
Vapor deposition could thus be expected to be the largest deep within the vent, and progressively reduce near the exit. 
This consideration suggests that the lower part of the vent is the first region that seals off.
The whole depth of the vent is not entirely filled by this process and sealing might happen while still leaving the upper portion of the vent unfilled.
However, in the present model, the vent height cancels out and our estimates for $\tau$ still hold.

As mentioned, our model takes the ice temperature to be uniform throughout the ice depth and constant during the plume expansion.
Two effects are to be mentioned:
\begin{enumerate}[i)]
  \item The ice temperature changes with depth, influencing the sticking coefficient;
  \item Water vapor, being warmer than the ice, results in an energy flux to the walls.
\end{enumerate}

Regarding the first point, one can expect the ice temperature to vary from about $273~\si{K}$ at the ice-liquid interface to $50$--$70~\si{K}$ at the surface, in the polar regions
of Europa \cite{rathbun2010galileo,ashkenazy2019surface}.
From \cite{carberry2022callisto}, one can see how the residence time of $\mathrm{H_2O}$ molecules impacting an ice layer changes dramatically with the ice temperature and 
the sticking coefficient changes accordingly.
For instance, Fig. C1 in \cite{carberry2022callisto} suggests a residence time in the order of a few seconds at an ice temperature of $165~\si{K}$.
This residence time increases exponentially as lower temperatures are considered, being approximately $10^{7}~\si{s}$ at a temperature of $110~\si{K}$, which is still 
much larger than the polar surface temperature of Europa.
At such cold temperatures, the sticking coefficient can be reasonably expected to be close to unity.
The implication for our model is that actual sealing might happen at an intermediate depth, where the ice temperature is sufficiently low for molecules to condense.
This might happen considerably below the surface.

In terms of energy deposition, the water vapor is expected to be warmer than the ice, resulting in an energy flux to the walls.
Besides, as vapor molecules deposit on the walls and accrete the ice, they release an energy that is related to the heat of sublimation.
Depending on the plume density and duration, these two factors might increase the local ice temperature, thus modifying the sticking coefficient and slowing down the vent-sealing 
process \cite{nakajima2016controlled}.
In certain conditions, at the largest temperatures, one might reach a balance between vapor deposition and sublimation, although numerical simulations suggest that the latter might be negligible \cite{van2024linking}.
For the considered reasons, we believe a complete analysis of this problem can only be done via depth-resolved numerical models.

Notice that capping or sealing, perhaps happening preferentially at some depth instead of the full height of the vent, might be compatible with unfilled crevices, such as the lineae on Europa, some of which could have depths sufficient to reach liquid sources.
Further mechanical considerations are possible, including an analysis of which cap thickness, $\mathrm{d}h$, is necessary to hold the pressure of an outgassing liquid. 
As for the other considerations, they go beyond the scope of the present analysis.

In consideration that vent geometry and sub-surface ice temperature during venting are not observationally constrained, we limit ourselves to the mentioned assumptions, which are sufficient for our order-of-magnitude estimates of the sealing time.


\section{Results and discussion}

\begin{figure}[htpb]
  \centering
  \includegraphics[width=1.0\textwidth]{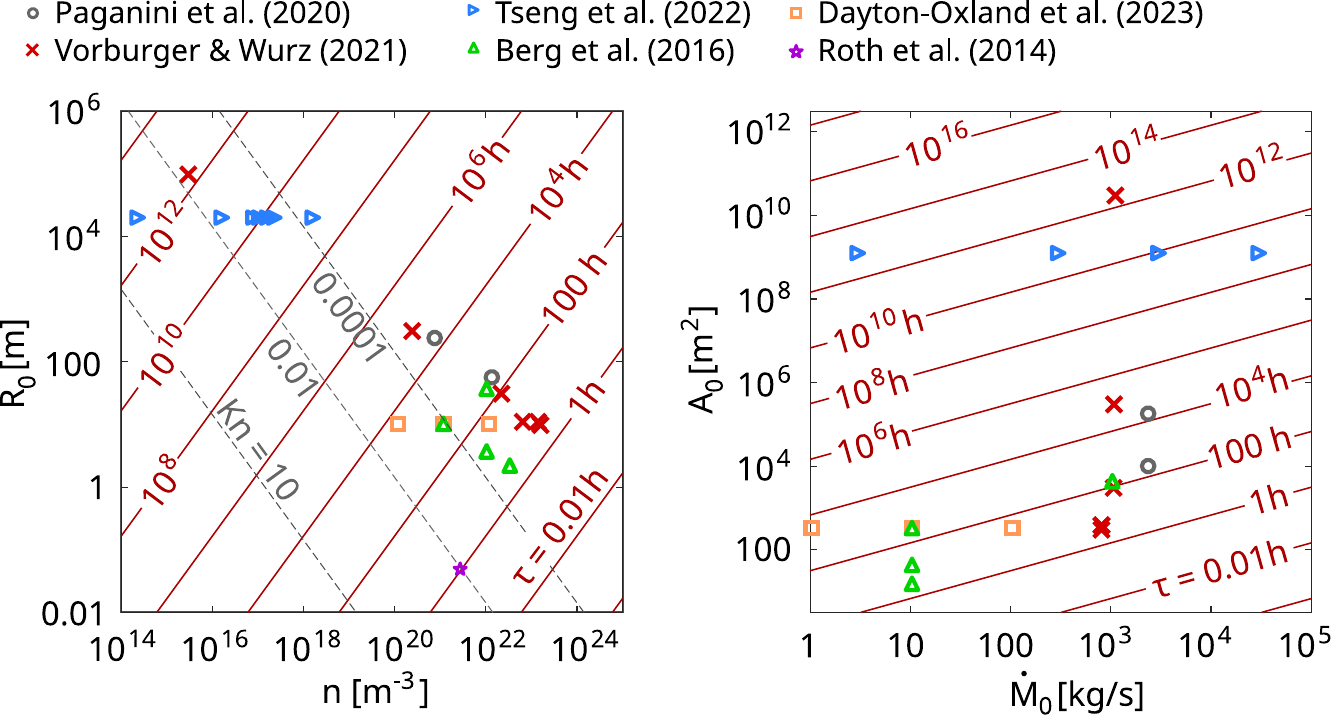}
  \caption{Cylindrical vents. Oblique red lines: contours of the vapor-deposition sealing time (Eq.~\eqref{eq:closing-time-cyl}), expressed in hours. Dashed grey lines: estimated Knudsen numbers. Symbols: selected values from the literature. Note: the point from \cite{roth2014transient}, is indicative only, as the authors consider a 2D aperture and not a cylindrical vent.} 
  \label{fig:results-literature-axi-plume}
\end{figure}

Figure~\ref{fig:results-literature-axi-plume} shows contours of $\tau_\mathrm{cyl}$, from Eq.~\eqref{eq:closing-time-cyl}, for different densities and initial vent radii (Fig.~\ref{fig:results-literature-axi-plume}-Left), or as a function of the initial aperture area and the initial mass-flow-rate (Fig.~\ref{fig:results-literature-axi-plume}-Right).
The contours are obtained with a temperature $T=273.16~\si{K}$, equal to the triple-point temperature of water, and an ice density $\rho_\mathrm{ice} = 917~\si{kg/m^3}$ \cite{ashkenazy2019surface}.
In the figure we explore a broad range of vapor densities.
Common estimates of the vapor density within the ice ridges are done based on the triple point pressure of water and on the ideal gas law, 
giving $n\approx1.6\times10^{23}~\si{m^{-3}}$, although this value might vary significantly along the deep crevices of Europa.

The sticking coefficient employed in Fig.~\ref{fig:results-literature-axi-plume} is $\alpha = 1$.
This is the physical upper bound and the dependence of $\tau$ with $\alpha^{-1}$ is linear:
a coefficient $\alpha=0.1$ would simply increase $\tau$ by a factor 10, shifting the contour lines rigidly in the logarithmic plot.
For sufficiently-dense plumes Fig.~\ref{fig:results-literature-axi-plume} suggests that vapor deposition can seal a sufficiently small vent in a matter of hours or minutes. 
For instance, a plume with density $n = 10^{22}~\si{m^{-3}}$ would fill a $R=1~\si{m}$ cylindrical aperture within about 6 hours, while an aperture of $R = 1~\si{cm}$ would be filled in about 1 minute.

\subsubsection*{Rarefaction regime}

Different points in Fig.~\ref{fig:results-literature-axi-plume}-Left correspond to different fluid-dynamic conditions \textit{inside} the vent.
For instance, low-density plumes flowing through small vents can be expected to be in the rarefied regime, showing large values of the Knudsen number, 
\begin{equation}\label{eq:definition-Kn-number}
  \mathrm{Kn} = \frac{\lambda}{R_0} = \frac{1}{\sqrt{2} \, n \sigma R_0} \, ,
\end{equation}
  
\noindent where we have selected the initial vent radius ($R_0$) as the characteristic dimension for the problem, and where $\lambda$ is the mean free path of the gas, based on an approximated hard-sphere elastic-collision cross-section $\sigma=5\times 10^{-19}~\si{m^2}$.
This value is taken as an approximation among the different values employed in the literature
\cite{crifo1989inferences,combi1996time,tenishev2008global,tucker20152d} and is sufficiently accurate for the sake of our analysis.

As mentioned, the vapor density within the ice ridges is typically estimated from the triple point of water, 
and this would result in a fluid-dynamically continuum regime at the spatial scales of interest.
Yet, in Fig.~\ref{fig:results-literature-axi-plume}, we also consider hypothetical low-density plumes.
Moreover, for any given density, as the vapor deposits to the walls and the aperture radius decreases, the Knudsen number progressively increases.
Eventually, just before the shutting-off time, gas-surface interactions are expected to be dominant over gas-phase molecular collisions.
In Fig.~\ref{fig:results-literature-axi-plume}-Left, we represent three contours of $\mathrm{Kn}$, as dashed lines. 
Below $\mathrm{Kn} = 0.0001$, the flow is in the continuum regime. 
For $0.01 \le \mathrm{Kn} \le 1$ (corresponding to lower densities and/or larger radii, as indicated in Eq.~\eqref{eq:definition-Kn-number}), the flow becomes transitional and is practically free-molecular around $\mathrm{Kn} = 10$.

The degree of rarefaction influences the validity of our model. 
We believe our vapor-deposition assumptions to be physically meaningful for relatively small Knudsen numbers, where the plume flowing through the vent can be interpreted as a collisional flow, and the deposition flux is indeed related to $\Gamma = \frac{1}{4}n v_\mathrm{th}$. 
In contrast, in a strongly-rarefied or collisionless plume expanding through a large vent, one might expect deposition to happen unevenly, 
following the angle of view instead of fluid-dynamic considerations.
The analysis of this latter case is suggested as a future work, to be explored via particle models such as DSMC or view-factor methods (see for instance \cite{araki2020radiosity}). 

Indicatively, we suggest $\mathrm{Kn} \approx 0.1$--$1$ as an upper limit for the validity of our model and we expect realistic vapor columns to be well below this threshold.
Full two or three-dimensional simulations would be necessary to fully establish the accuracy of our model, that is also expected to depend on the specific geometry to be considered, such as a nozzle-shaped vent, constant-section cylinder, or an irregular aperture (e.g. \cite{tucker20152d} and \cite{van2024linking}).
Nevertheless, our suggestion is sufficient for our order-of-magnitude analysis. 
Finally, we stress that that our analysis and the Knudsen number contours shown in Fig.~\ref{fig:results-literature-axi-plume} are strictly limited to the vent region and are not representative of the rarefaction occurring in the further planetary-scale plume expansion.

\subsubsection*{The 2012 HST observation}

In 2012, the Hubble Space Telescope (HST) detected consistent plume activity over a 7-hour period \cite{roth2014transient}.
Notice that this observation window constitutes a lower bound for the plume activity, which could have lasted longer.
Assuming the plume originated from an elongated surfacial-ice fracture and taking an ice temperature of $230~\si{K}$, the authors obtained a density $n\approx 3\times 10^{21}~\si{m^{-3}}$, and the vent was estimated to be a few cm thick (see the supplemental material of the same article). 
Introducing these values in our two-dimensional sealing time expression, Eq.~\eqref{eq:closing-time-2D}, considering $D_0 = 5~\si{cm}$, we predict that vapor deposition alone would seal a fracture in $\tau_\mathrm{2D} = 0.58 ~\si{h}$.
This prediction differs from the observed 7-hour plume by about a factor 10 (or more if the plume did last longer than the observation).
We believe this could be explained by
\begin{itemize}
  \item The fracture being wider and/or the plume being less dense than initially suggested in \cite{roth2014transient}: 
  from Eq.~\eqref{eq:closing-time-2D}, a ratio $D_0/n$ about 10 times larger would let a plume be active throughout a 7-hour observation window.
  \item The approximate nature of our model and the uncertainty in the physical parameters employed by both the authors and us (consider for instance the vapor temperature and the sticking coefficient, $\alpha$, that we simply set to 1, while a lower value would increase the predicted sealing time);
\end{itemize}

\noindent Ultimately, our order-of-magnitude analysis suggests that water-vapor deposition is an important phenomenon at these plume conditions, 
the associated time scales being roughly in-line with the known plume duration.
In other words, it is physically conceivable that the 2012 plume was shut off entirely by vapor deposition at the walls. 
This is not conclusive evidence, since the 2012 plume might have lasted long past the observation window. 


\subsubsection*{Other references}

In Fig.~\ref{fig:results-literature-axi-plume}, we include the density and vent-size values employed in selected literature references.
These references consider various plume cases, either observed or simulated numerically, and each case is represented with a separate symbol in the figure.
We have excluded from our analysis a number of references, that consider point-source plumes \cite{teolis2017plume,jia2018evidence,arnold2019magnetic,huybrighs2020active}.
These references cannot be plotted in Fig.~\ref{fig:results-literature-axi-plume}, as point-source vents would have a radius $R\rightarrow0$ and would seal instantaneously in our model ($\tau \rightarrow 0$).
Yet, our sealing-time equations can be applied to such cases and would allow to infer the minimum vent size that allows for a sufficiently long-lived plume.
The references included in Fig.~\ref{fig:results-literature-axi-plume} are:
\begin{description}
  \item[Paganini et al. (2020) \cite{paganini2020measurement}] report a detection from the Keck Observatory, occurred on April 26th, 2016, compatible with a $2360~\si{kg/s}$ plume.
  The two points shown in Fig.~\ref{fig:results-literature-axi-plume} represent the extremes of a range of possible vent sizes, as suggested by the authors.
  \item[Vorburger \& Wurz (2021) \cite{vorburger2021modeling}] perform Monte Carlo simulations of different sources, representing a near-surface liquid inclusion, a diapir and an ocean plume, with various conditions; 
  \item[Tseng et al. (2022) \cite{tseng20223d}] perform collisional DSMC simulations spanning a large range of mass flow rates and relatively large areas;
  \item[Berg et al. (2016) \cite{berg2016dsmc}] also perform collisional DSMC simulations, but consider higher densities than Tseng et al. (2022) \cite{tseng20223d}, 
        and smaller vent radii;
  \item[Dayton-Oxland et al. (2023) \cite{dayton2023situ}] elaborate on the simulations by Berg et al. (2016) \cite{berg2016dsmc} aiming to give predictions useful to the Juice mission. 
\end{description}

The vent area, the plume density and the mass flow rate from said references are also reported in Table~\ref{tab:parameters-from-literature}.
All such references refer to cylindrical vents and the sealing-time contours are obtained from Eq.~\ref{eq:closing-time-cyl}.
Nevertheless, we believe it is illustrative to include in the figure the two-dimensional case of \cite{roth2014transient}.
Indeed, employing the thickness width $D_0$ instead of the radius in the cylindrical formula gives a sealing time of about $1~\si{h}$, that is within a factor 2 of the two-dimensional result.
We report this case only in Fig.~\ref{fig:results-literature-axi-plume}-Left and exclude it from Fig.~\ref{fig:results-literature-axi-plume}-Right, since a sealing-time prediction based on the areas and the mass-flow-rate would instead differ vastly between the cylindrical and 2D cases.
This is due to a fundamental difference between these models, as in the 2D case one could have large mass flow rates caused by narrow but elongated cracks,
while in the case of axisymmetric vents a large mass flow rate would necessarily imply a large radius.

Figure~\ref{fig:results-literature-axi-plume} also suggests that, if a sufficiently large area is assumed (consider for instance the simulations in \cite{tseng20223d} or the large-radius case of \cite{vorburger2021modeling}), then aperture sealing by vapor deposition is negligible. 
In such cases other physical mechanisms---such as water spillage and freezing or geological activity  \cite{vincent2022numerical,quick2013constraints}---are necessary to shut off a plume.
Otherwise, according to the vent sizes implemented in the aforementioned models, the plumes would be long-lived, on the order of 100s of hours or more, 
and should be readily detectable, which is inconsistent with recent attempts at observing such plumes.

We stress that our observations do not necessarily impact on the validity of the referenced simulation works.
Indeed, combinations of smaller densities and larger areas (provided that the mass flow rate is preserved) are often employed for numerical reasons and 
do not affect the overall external plume evolution, that is the target of the studies.
Nonetheless, we believe it is useful to put such simulation conditions in the context of this work.

\begin{table}[htpb]
  \centering
  \begin{tabular}{l|c|c|c}
    Ref. & $A_0~\si{[m^2]}$ & $n~\si{[m^{-3}]}$ & $\dot{M}_0~\si{[kg/s]}$ \\
    \hline
    Paganini et al. (2020) \cite{paganini2020measurement} & $10^4$ & $1.35\times10^{22}$ & $2360$ \\
    Paganini et al. (2020) \cite{paganini2020measurement} & $18\times10^{4}$ & $7.52\times10^{20}$ & $2360$ \\
    Vorburger \& Wurz (2021) \cite{vorburger2021modeling}   &  $3.0\times10^{2}$    &    $1.59\times10^{23}$  &    $807.3$  \\ 
    Vorburger \& Wurz (2021) \cite{vorburger2021modeling}   &  $3.0\times10^{10}$   &    $2.95\times10^{15}$  &    $1112.28$  \\ 
    Vorburger \& Wurz (2021) \cite{vorburger2021modeling}   &  $3.0\times10^{5}$    &    $2.41\times10^{20}$  &    $1076.4$  \\
    Vorburger \& Wurz (2021) \cite{vorburger2021modeling}   &  $3.0\times10^{3}$    &    $2.20\times10^{22}$  &    $1067.43$  \\
    Vorburger \& Wurz (2021) \cite{vorburger2021modeling}   &  $3.8\times10^{2}$    &    $1.40\times10^{23}$  &    $807.84$  \\
    Vorburger \& Wurz (2021) \cite{vorburger2021modeling}   &  $3.8\times10^{2}$    &    $6.47\times10^{22}$  &    $807.84$  \\
    Tseng et al. (2022) \cite{tseng20223d}             & $1.26\times10^9$     & $2.27\times10^{14}$   & $2.99\times10^{0}$ \\
    Tseng et al. (2022) \cite{tseng20223d}             & $1.26\times10^9$     & $1.59\times10^{18}$   & $2.99\times10^{4}$ \\
    Tseng et al. (2022) \cite{tseng20223d}             & $1.26\times10^9$     & $1.59\times10^{17}$   & $2.99\times10^{3}$ \\
    Tseng et al. (2022) \cite{tseng20223d}             & $1.26\times10^9$     & $1.59\times10^{16}$   & $2.99\times10^{2}$ \\
    Tseng et al. (2022) \cite{tseng20223d}             & $1.26\times10^9$     & $2.27\times10^{17}$   & $2.99\times10^{3}$ \\
    Tseng et al. (2022) \cite{tseng20223d}             & $1.26\times10^9$     & $1.06\times10^{17}$   & $2.99\times10^{3}$ \\
    Tseng et al. (2022) \cite{tseng20223d}             & $1.26\times10^9$     & $7.96\times10^{16}$   & $2.99\times10^{3}$ \\ 
    Berg et al. (2016) \cite{berg2016dsmc}            &    $1.52\times10^1$ &  $3.35\times10^{22}$   &  $1.04\times10^1$ \\
    Berg et al. (2016) \cite{berg2016dsmc}            &    $4.3 \times10^1$ &  $1.04\times10^{22}$   &  $1.04\times10^1$ \\
    Berg et al. (2016) \cite{berg2016dsmc}            &    $3.27\times10^2$ &  $1.17\times10^{21}$   &  $1.04\times10^1$ \\
    Berg et al. (2016) \cite{berg2016dsmc}            &    $4.3 \times10^3$ &  $1.04\times10^{22}$   &  $1.04\times10^3$ \\
    Dayton-Oxland et al. (2023) \cite{dayton2023situ}          & $3.27\times10^2$  &   $1.18\times10^{20}$   &  $1.04\times10^0$ \\ 
    Dayton-Oxland et al. (2023) \cite{dayton2023situ}          & $3.27\times10^2$  &   $1.18\times10^{21}$   &  $1.04\times10^1$ \\
    Dayton-Oxland et al. (2023) \cite{dayton2023situ}          & $3.27\times10^2$  &   $1.18\times10^{22}$   &  $1.04\times10^2$ \\
  \end{tabular}
  \caption{Parameters employed by the considered references, shown as symbols in Fig.~\ref{fig:results-literature-axi-plume}. Cylindrical vents.} 
  \label{tab:parameters-from-literature}
\end{table}

 
\section{Conclusion}

In this work, we have built a model to estimate the rate at which vapor deposition would close off a vent in Europa's ice crust.
We consider the shutting-off times of a plume originating in the sub-surface, as a function of its density, mass flow rate and of the aperture size.
Our model does not consider water spillage, tides, geological activity or non-uniformities in the ice temperature and physical properties.
Also, we do not consider energy transfer to the walls, caused by deposited molecules.
This effect might modify the local ice temperature, affecting the sticking coefficient and ultimately the sealing times.

Both two-dimensional elongated-fracture vents and cylindrical apertures are considered.
Unless tidal forces keep a fracture open, then our model offers an upper bound for the duration of a plume originating from a vent. 
If an aperture is too small, for a given plume density, then vapor deposition would seal it within minutes or hours.
As an implication to numerical simulations, models based on vents that are too small might be physically unrealistic and incompatible with sustained plumes. 
Conversely, vents that are excessively large cannot be realistically sealed by vapor deposition and require other physical mechanisms to be considered.

When applied to the proposed fracture conditions associated with the 2012 Hubble Space Telescope plume detection \cite{roth2014transient} 
our model results in a sealing time of about 30 minutes. 
In contrast, the plume was continually observed for 7 hours and might have lasted longer. 
While we only target an order-of-magnitude analysis, our prediction suggests that the actual fracture size might have been larger than 
initially proposed by \cite{roth2014transient}, and/or the plume density at the vent might have been lower.
Yet, within the limitations of our simple model, the fast predicted sealing times indicate that water-vapor deposition is an important phenomenon at the considered conditions
and might potentially be responsible for entirely shutting off a plume.

As mentioned, our estimates neglect important effects such as continued tidal forces that might cause the surface fractures to remain open (as in the case of the Tiger Stripes on Enceladus \cite{rhoden2020formation}).
However, such effects could be expected to be less relevant on Europa, due to its larger size and/or thicker ice shell, making our model of interest to future simulation activities and to the interpretation of observations.
Such considerations go beyond the scope of this work.


\section*{Acknowledgments}

SB’s research was supported by an appointment to the NASA Postdoctoral Program at NASA Goddard Space Flight Center, administered by Oak Ridge Associated Universities under contract with NASA.
SRCM's research was supported by a Solar System Workings grant, number 80NSSC22K0097.
OJT's research was supported by a Solar System Workings grant, number 80NSSC24K0657.

\end{document}